\documentclass{caosp309}

\usepackage{graphicx}

\usepackage{natbib}
\newcommand{\MS}{\ifmmode{\,}\else\thinspace\fi{\rm M}\ifmmode_{\odot}\else$_{\odot}$\fi}
\articleNo{301}
\pubyear{2020}
\volume{50}
\volnumber{3}
\firstpage{1}
\received{May 1, 2020}
\accepted{July 28, 2020}

\def\BibTeX{{\rm B\kern-.05em{\sc i\kern-.025em b}\kern-.08em
             T\kern-.1667em\lower.7ex\hbox{E}\kern-.125emX}}

\begin{document}

%
\hauthor{R. S. Rathour, et al.}

\title{Non-evolutionary effects on period change in Magellanic Cepheids}


%
%
\author{
        R. S. Rathour\inst{1}\orcid{0000-0002-7448-4285}
      \and
        R. Smolec\inst{1}\orcid{0000-0001-7217-4884} 
      \and 
        G. Hajdu\inst{1}\orcid{0000-0003-0594-9138}   
      \and 
        P. Karczmarek\inst{2}\orcid{0000-0002-0136-0046}
      \and \\
        V. Hocd\'e\inst{1}\orcid{0000-0002-3643-0366}
      \and 
        O. Zi\'ołkowska\inst{1}\orcid{0000-0002-0696-2839}
      \and 
        I. Soszy\'nski\inst{3}\orcid{0000-0002-7777-0842}
      \and 
         A. Udalski\inst{3}\orcid{0000-0001-5207-5619}
      }

%
\institute{
           Nicolaus Copernicus Astronomical Centre, Polish Academy of Sciences, Bartycka 18, 00-716 Warszawa, Poland, \email{\texttt{rajeevsr@camk.edu.pl}}
         \and 
          Departamento de Astronomía, Universidad de Concepción, Casilla 160-C, Concepción, Chile
         \and 
          Astronomical Observatory, University of Warsaw, Aleje Ujazdowskie 4, Warsaw, 00-478, Poland
          }

\date{March 8, 2003}

\maketitle

\begin{abstract}

Classical Cepheids are a cornerstone class of pulsators, fundamental to testing stellar evolution and pulsation theories. Their secular period changes, characterized through 
$O-C$ (Observed minus Calculated) diagrams, offer valuable insights into their evolution. While evolutionary period changes are well understood from both observational and theoretical perspectives, shorter timescale period changes (on the order of 
($\sim$ 10$^{2}$-10$^{4}$ days) - known as non-evolutionary period changes are yet to be systematically explored.

In this work, we present a detailed and comprehensive search for non-evolutionary period changes using $O-C$ analysis of Magellanic Cloud (MC) Cepheids, based on 20+ years of OGLE photometry data. Our sample includes both the Large Magellanic Cloud (LMC) and the Small Magellanic Cloud (SMC) Cepheids, focusing on single radial mode Cepheids (both fundamental (FU) and first overtone (FO) modes). The results are grouped into two phenomena: (a) Cepheids in binary systems (b) Non-linear period changes.

\end{abstract}

%
\section{Introduction}
\label{Introduction}

Classical Cepheids (hereafter Cepheids) are intermediate- to high-mass ($3-13\MS$) supergiant stars undergoing core-helium burning. These variable stars are known for their large-amplitude radial pulsations and are found within a narrow region in the Hertzsprung-Russell diagram known as the classical instability strip (IS). Cepheids hold a pivotal role as standard candles due to their well-defined period-luminosity (P-L) relation, also referred to as the Leavitt law \citep{Leavitt1912HarCi.173....1L}. As these stars evolve across the IS, Cepheids experience changes in their pulsation periods as a consequence of change in their internal structure. The $O-C$ diagram technique provides a valuable perspective on both the evolutionary and non-evolutionary aspects affecting these pulsating stars.

Extensive investigations have been conducted on evolutionary period changes in Cepheids \citep[e.g.,][]{PietrukowiczLMC2001AcA....51..247P, PietrukowiczSMC2002AcA....52..177P, Turner2006PASP..118..410T, Karczmarek2011AcA....61..303K, Rodriguez-Segovia2022MNRAS.509.2885R}. In contrast, shorter-timescale period changes ($\sim$ 10$^{2}$-10$^{4}$ days) arising from non-evolutionary origins remain largely unexplored \citep[e.g.,][]{Poleski2008AcA....58..313P}. Notably, the Cepheid V1154 Cygni is studied in detail for cycle-to-cycle period fluctuations changes using \textit{Kepler} data, revealing light curve modulation and granulation signatures \citep{Derekas2012MNRAS.425.1312D,Derekas2017MNRAS.464.1553D}.
Among these non-secular period changes, a fraction of them which are periodic in nature are due to light-travel time effect \citep[LTTE;][]{Irwin1952ApJ...116..211I, Irwin1959AJ.....64..149I}, suggesting that these Cepheids may belong to binary systems. In the Milky Way (MW), a high binary fraction among Cepheids (up to 80\%) is expected \citep{Szabados2003ASPC..298..237S, Evans1992ApJ...384..220E, Evans2013AJ....146...93E}, and out of $\sim$15,000 known MW Cepheids \citep{Pietrukowicz2021AcA....71..205P}, only $\sim$200-300 have been spectroscopically confirmed as binaries \citep[see][]{Szabados2003bIBVS.5394....1S, Anderson2024A&A...686A.177A}. In the Magellanic Clouds, the numbers are even lower, with only $\sim$14 confirmed in the SMC and $\sim$27 in the LMC \citep{Szabados1983Ap&SS..96..185S, Pilecki2021ApJ...910..118P, Pilecki2024A&A...686A.263P}. This gap highlights the need for a comprehensive study of period changes in MC Cepheids.

We use of OGLE photometry data spanning over 20 years, specifically the I-band MC data from the OGLE-III and OGLE-IV phases \citep[and OGLE references therein]{Soszynski2019AcA....69...87S} of the survey. To construct $O-C$ diagrams, we employ the modified Hertzsprung method \citep[e.g.,][]{Hajdu2021ApJ...915...50H}. These $O-C$ diagrams are then categorized by shape into three primary groups: linear, parabolic, and others. The third group, consisting of non-linear period change candidates, are further divided into periodic and aperiodic (irregular) $O-C$ groups.

\section{Periodic $O-C$: binary candidates}
\label{Periodic}
\begin{figure}[ht]
\centering
\includegraphics[width=0.49\textwidth]{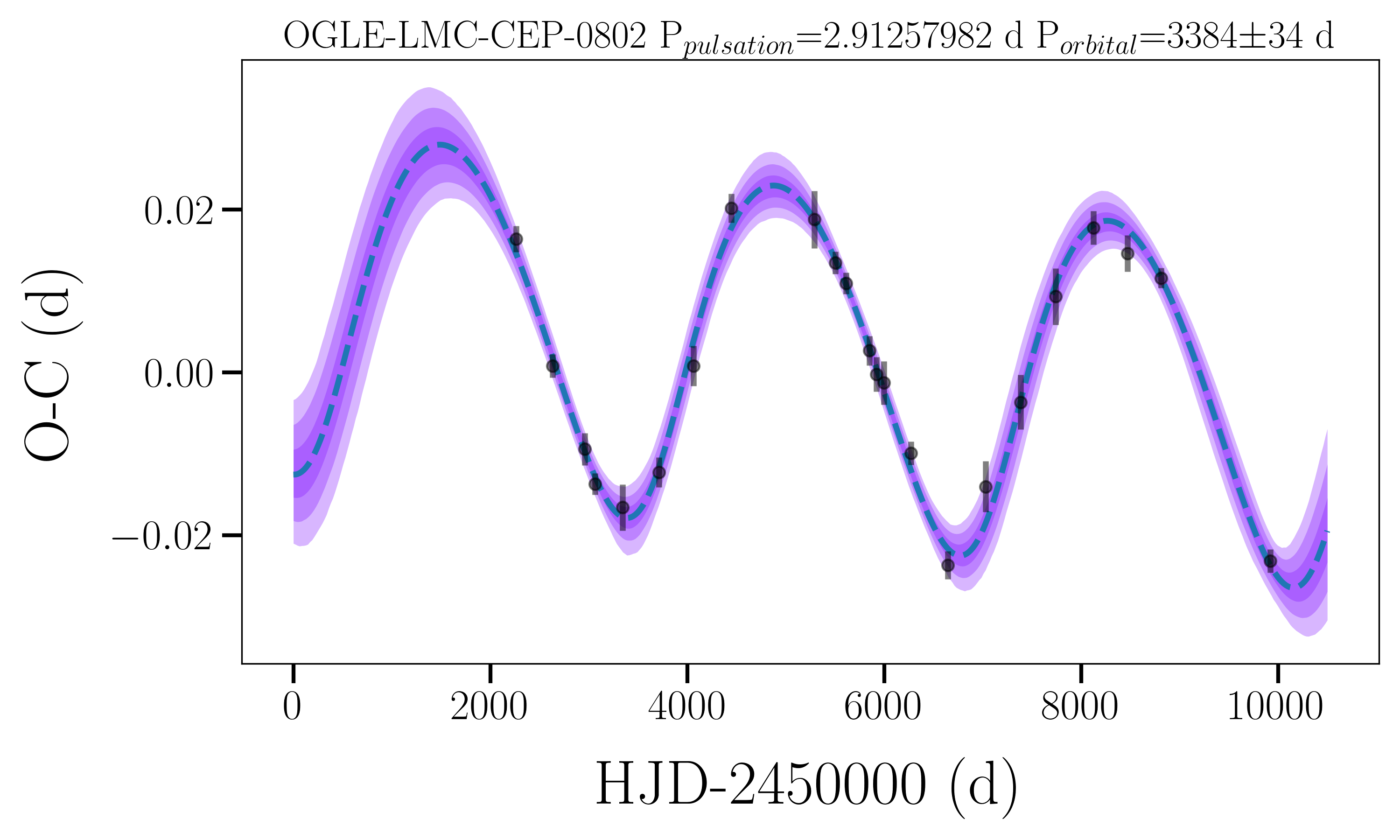}
\includegraphics[width=0.49\textwidth]{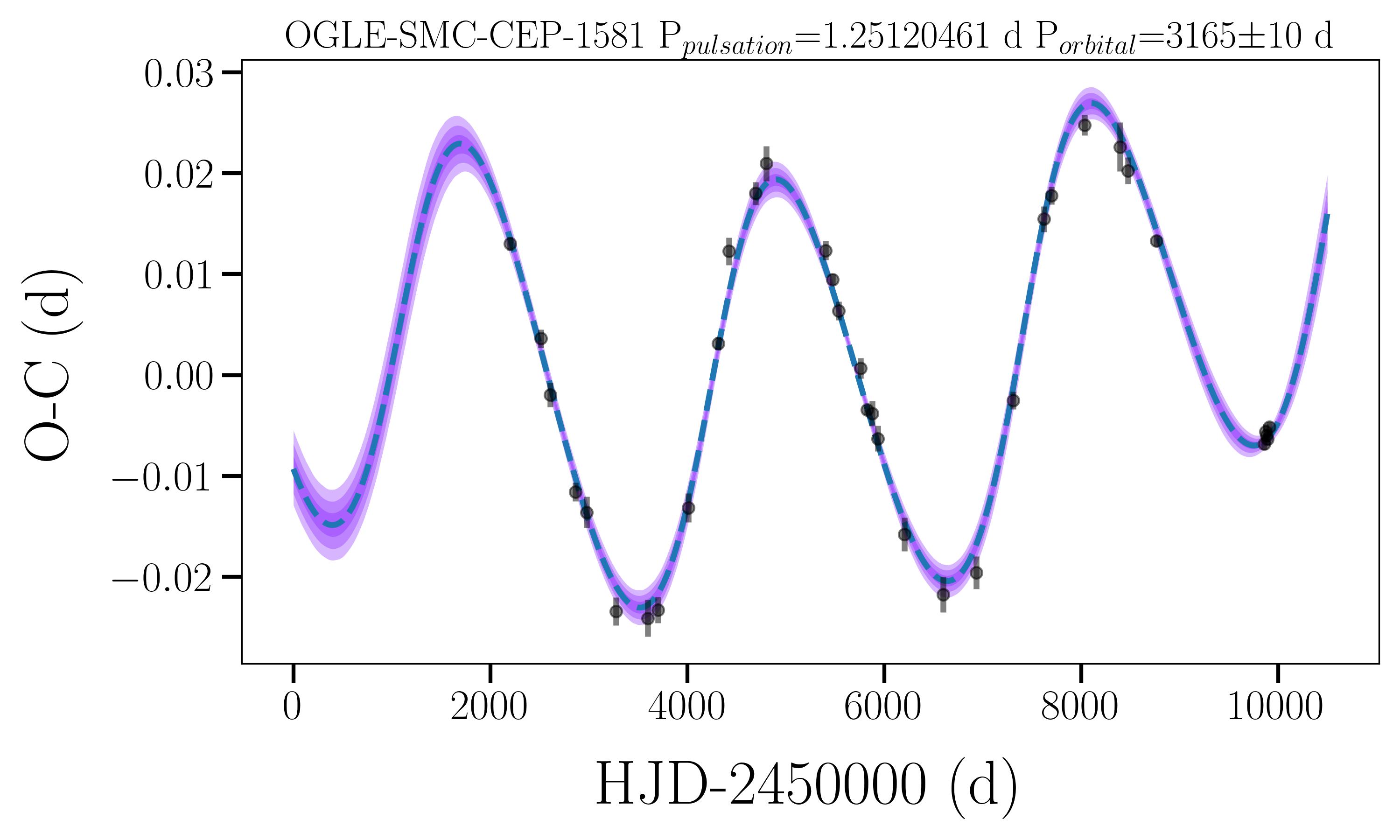}
\caption{Example $O-C$ diagram of FU mode candidates in LMC (\textit{left}) and SMC (\textit{right}). The dashed blue line depicts the median MCMC binary fit solution along with shaded regions showing uncertainty the fit. The header shows its OGLE-ID, pulsation period (in days) and orbital period (in days)}
\label{fig:binary_O-C_diagrams}
\end{figure}

\begin{figure}[ht]
\centering
\includegraphics[height=5cm, width=0.98\textwidth]{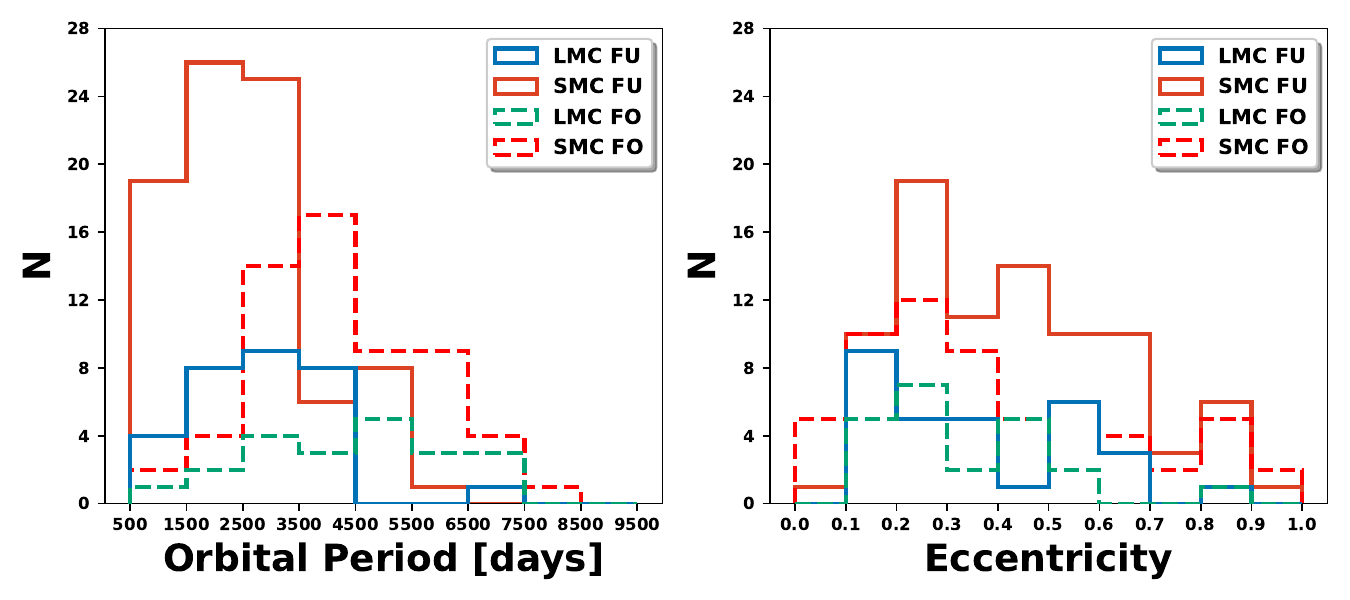}
\caption{Distribution of orbital periods (\textit{left}) and eccentricities (\textit{right}) for binary Cepheid candidates.}
\label{fig:binary_distribution}
\end{figure}

The analysis of $O-C$ diagrams followed by filtering candidates with periodic signatures were subjected to our Bayesian binary fitting routine. In total this resulted in 197 Cepheids candidates (52 in LMC and 145 in SMC) which show non-evolutionary period changes associated with binarity. In Fig.~\ref{fig:binary_O-C_diagrams} some example $O-C$ diagrams are shown. For all final candidates, we determined a total of five orbital parameters: orbital period ($P_\mathrm{orb}$), time of periastron passage ($T_0$), eccentricity ($e$), argument of the periastron ($\omega$), projected semi-major axis ($a\sin i$). Using these, we also calculate three more derived parameters, period change rate ($\beta$), radial velocity semi-amplitude ($K$) and mass function ($f(m)$). The distribution of important binary parameters are presented in Fig.~\ref{fig:binary_distribution}.

Cepheids that are systematically brighter than the Period-Luminosity relation were investigated by \cite{Pilecki2021ApJ...910..118P} to search for binarity. Their work hypothesizes that such Cepheids are brighter because of the additional light contribution from the companion likely to be in a giant phase. We searched for such `\textit{overbright}' Cepheids in our sample and obtained four such candidates (two in each Magellanic Cloud). The LMC candidates OGLE-LMC-CEP-0837 and OGLE-LMC-CEP-0889 were previously identified by \cite{Pilecki2021ApJ...910..118P} as SB2-type systems. Whereas the SMC candidates OGLE-SMC-CEP-3061 and OGLE-SMC-CEP-4365 are new detections and are yet to be confirmed spectroscopically. For details of the work refer to \cite{Rathour2024A&A...686A.268R}.

\section{Irregular $O-C$: non-linear period change candidates}
\label{Irregular}
\begin{figure}[ht]
\centering
\includegraphics[width=0.49\textwidth]{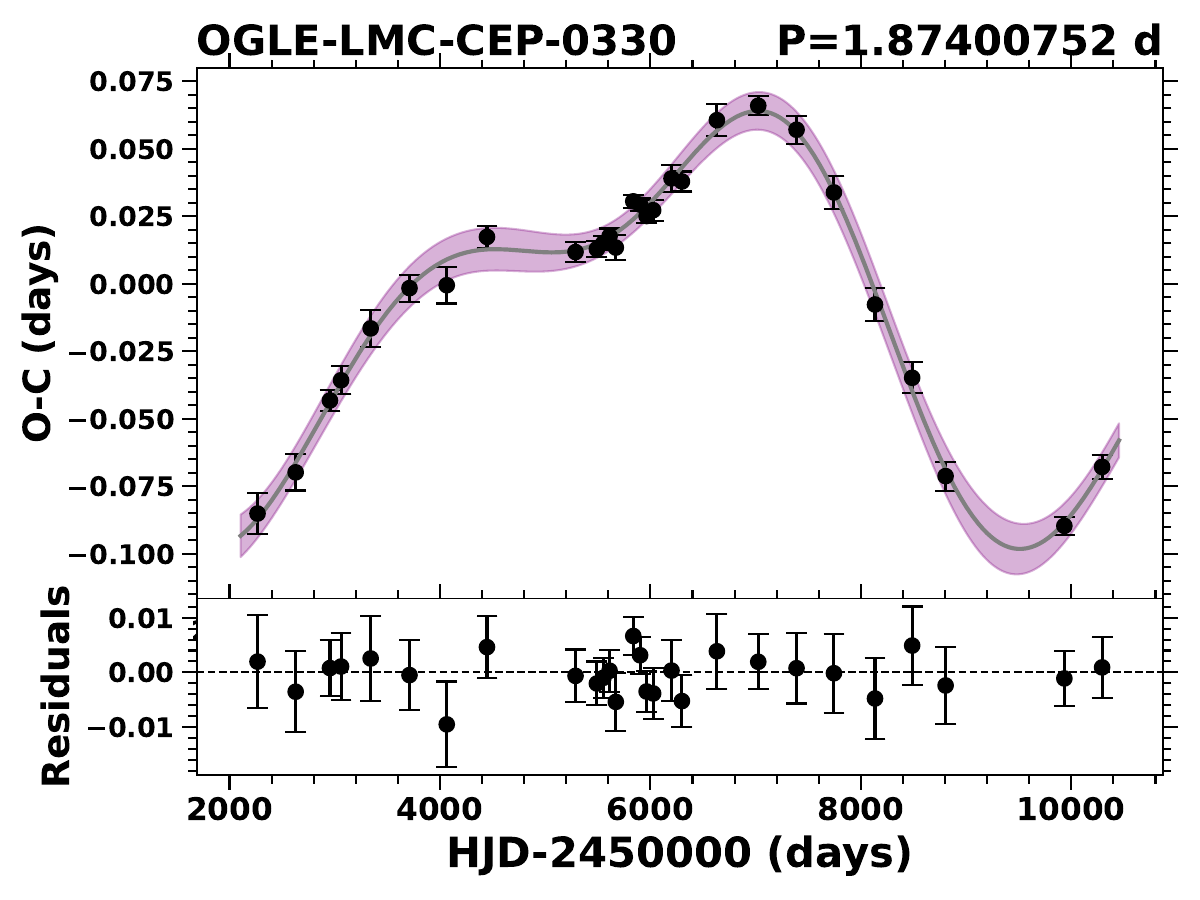}
\includegraphics[width=0.49\textwidth]{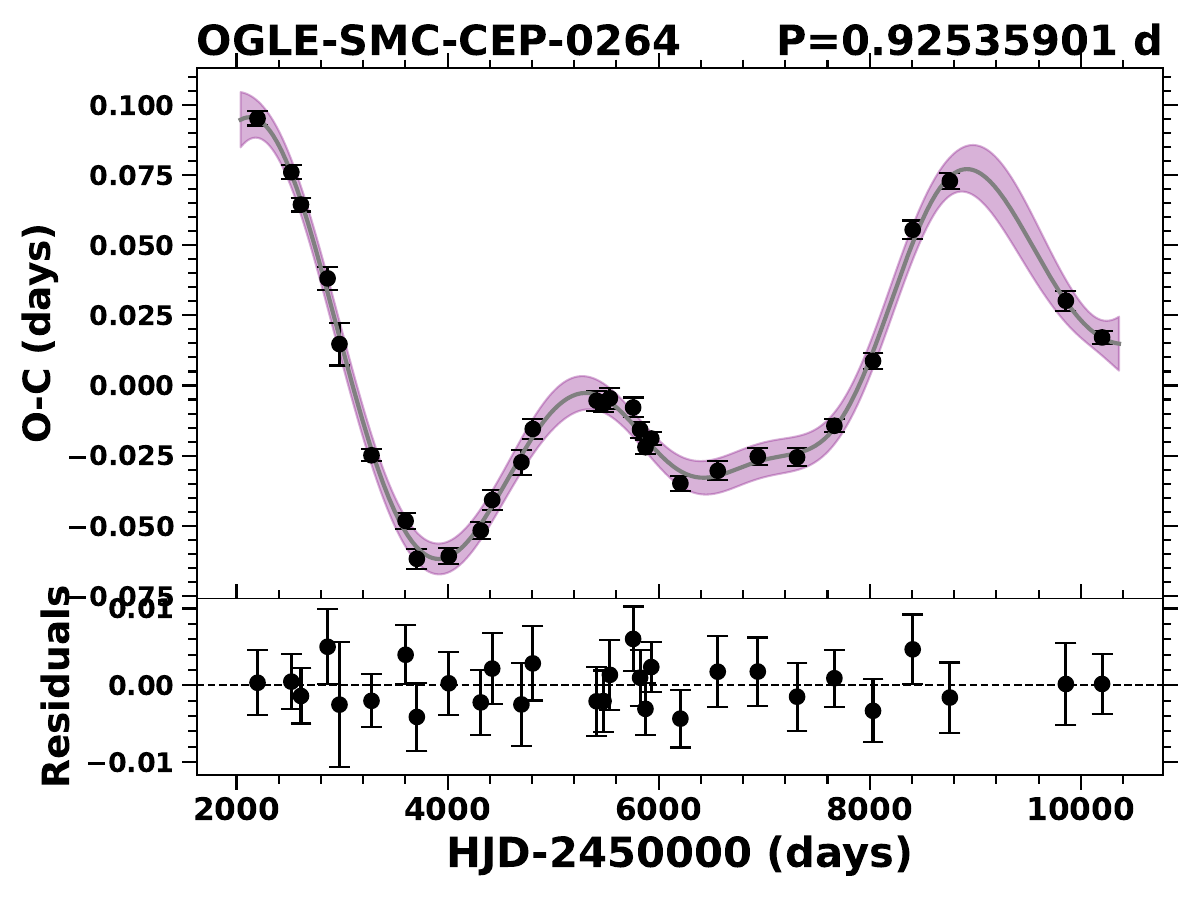}
\caption{Example $O-C$ diagrams of FO mode irregular period change candidates in LMC (\textit{left}) and the SMC (\textit{right}) candidates. The header shows the OGLE-ID and pulsation period (in days).}
\label{fig:irregular_O-C_diagrams}
\end{figure}

In a sample of 1,585 Cepheids, we identified non-linear period changes exhibiting varied behaviors, ranging from quasi-periodic to abrupt and complex $O-C$ patterns, as illustrated by examples in Fig.~\ref{fig:irregular_O-C_diagrams}. The origin of these irregular period changes remains unknown, and no definitive mechanism has been proposed to explain them. To characterize these candidates, we employ several techniques, including the Eddington–Plakidis test \citep{Eddington1929MNRAS..90...65E}, time-frequency analysis, instantaneous period method, and the Stetson $L$ index \citep{Stetson1996PASP..108..851S}.

\begin{figure}[ht]
\centering
\includegraphics[height=4.5cm, width=0.98\textwidth]{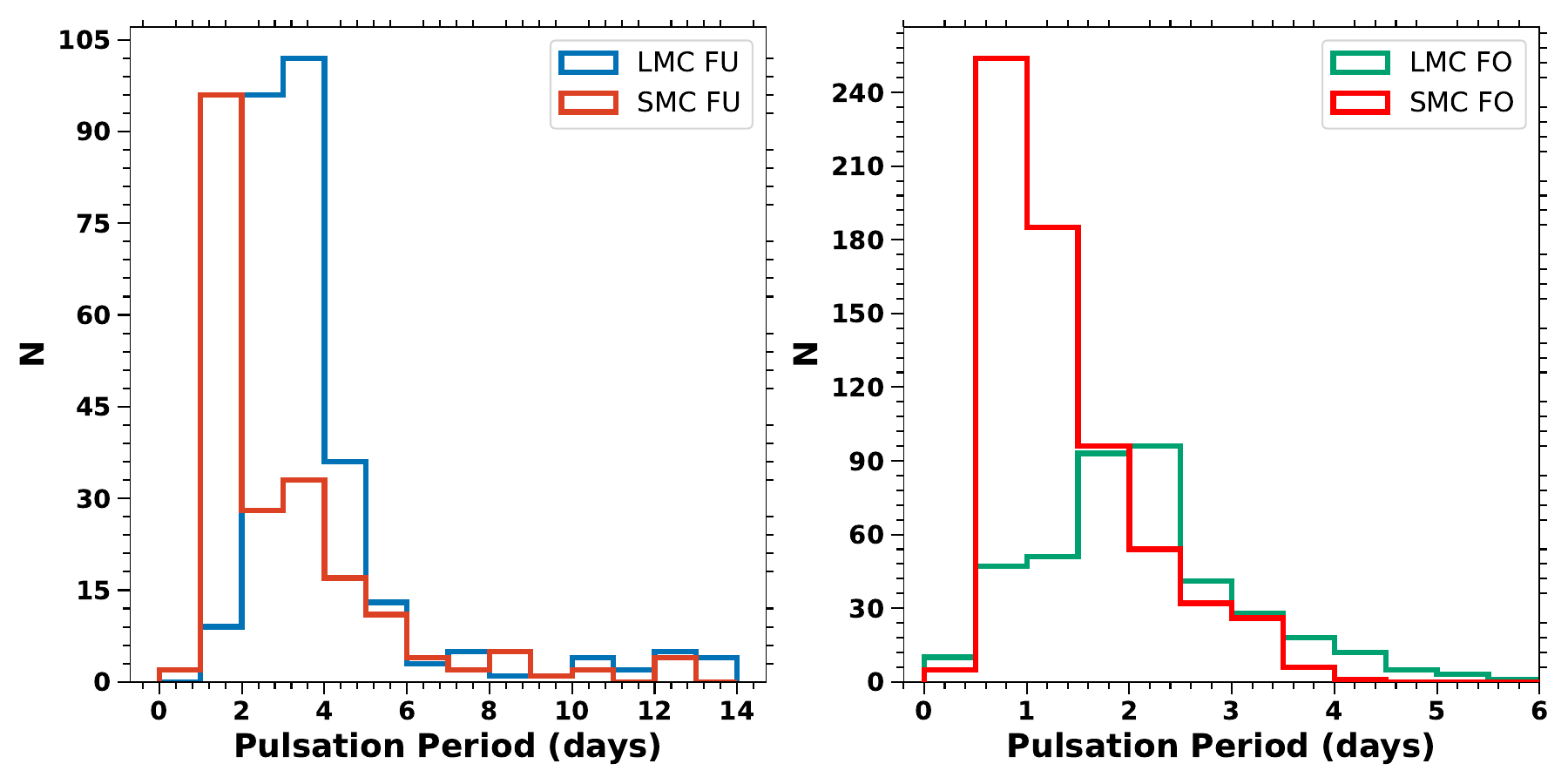}
\caption{Pulsation period distribution of irregular period change candidates.}
\label{fig:irregular_distribution}
\end{figure}

The distribution of these irregular period change candidates as a function of pulsation period is shown in Fig.~\ref{fig:irregular_distribution}. Detailed incidence rates including characterization of their amplitudes and timescales, will be presented in our upcoming work (Rathour et al. \textit{in prep.}).

\section{Conclusions}
\label{Conclusions}

Our comprehensive search for non-evolutionary period changes using $O-C$ analysis of Cepheids from OGLE photometry has resulted in a compilation of two classes of candidates. The first category contains 197 candidates showing light travel time effect signature meaning they are Cepheid candidates likely of binarity nature: (LMC: 30 (FU) and 22 (FO); SMC: 85 (FU) and 60 (FO)). Theses candidates indicate a detection fraction of binary Cepheid candidates: LMC: 1.7 \% (FU); 1.8 \% (FO); SMC: 3.3 \% (FU); SMC 3.7 \% (FO). The second class consists of a sample of 1585 Cepheids candidates exhibiting irregularities in their $O-C$ diagrams corresponding to non-evolutionary period changes of unknown origin. The inventory contains 695 LMC Cepheids (FU: 290 and FO: 405) and 890 SMC Cepheids (FU: 231 and FO: 659). With FO mode Cepheids exhibiting irregular period changes in systematically larger fraction of the sample than FU mode ones, we conclude that FO mode is more susceptible to period change, which in agreement with previous findings by \cite{Poleski2008AcA....58..313P}.


\acknowledgements
RSR, RS, and OZ are supported by the National Science Center, Poland, Sonata BIS project 2018/30/E/ST9/00598. GH, VH and PK acknowledges grant support from European Research Council (ERC) under the European Union's Horizon 2020 research and innovation program (grant agreement No. 695099). IS and AU are funded by the National Science Centre, Poland, grant no. 2022/45/B/ST9/00243.

\bibliography{demo_caosp309}

\clearpage

\end{document}